\documentclass[prb,aps,showpacs,twocolumn,unsortedaddress]{revtex4}

\usepackage{graphics, bm, xspace}
\usepackage{graphicx}
\usepackage{psfrag}
\usepackage{amsmath}
\usepackage{amssymb}
\usepackage{epsfig}
\usepackage{subfigure}
\usepackage{grffile}
\usepackage{times}
\usepackage{color}
\usepackage{multirow}
\usepackage[bookmarks=false,linkcolor=blue,urlcolor=blue,colorlinks,citecolor=blue]{hyperref}
\usepackage{float}
\usepackage{gensymb}

\newcommand{\ang}    {\mbox{\AA}}
\newcommand{\beq}    {\begin{equation}}
\newcommand{\enq}    {\end{equation}}
\newcommand{\ceq}[1] {(\ref{#1})}

\newcommand{\aav}     {{\bf a}}
\newcommand{\dd}     {{\bf d}}
\newcommand{\kk}     {{\bf k}}

\newcommand{\qq}     {{\bf q}}

\newcommand{\xx}     {{\bf x}}
\newcommand{\yy}     {{\bf y}}

\newcommand{\KK}     {{\bf K}}
\newcommand{\pp}     {{\bf p}}

\newcommand{\GG}     {{\bf G}}

\newcommand{\df}     {\equiv}

\newcommand{\mos}    {${\rm MoS_2}$\xspace}
\newcommand{\mose}   {${\rm MoSe_2}$\xspace}

\newcommand{\tase}   {${\rm TaSe_2}$\xspace}
\newcommand{\tas}    {${\rm TaS_2}$\xspace}

\newcommand{\nbse}   {${\rm NbSe_2}$\xspace}
\newcommand{\nbs}    {${\rm NbS_2}$\xspace}

\newcommand{\stau}     {{\boldsymbol\tau}}

\newcommand{\deltaind}{$\Delta_{\rm ind}$\xspace}

\newcommand{\deltag}{$\Delta_\Gamma$\xspace}

\begin{document}

\title{Superconductivity in twisted Graphene \nbse heterostructures}
\author{Yohanes S. Gani$^1$, Hadar Steinberg$^2$,  Enrico Rossi$^1$}
\affiliation{
             $^1$Department of Physics, William \& Mary, Williamsburg, VA 23187, USA,\\
             $^2$The Racah Institute of Physics, The Hebrew University of Jerusalem, Jerusalem 91904, Israel
            }
\date{\today}

   
\begin{abstract}
We study the low-energy electronic structure of heterostructures formed by one sheet of graphene placed on 
a monolayer of \nbse. We build a continuous low-energy effective model that takes into account the presence
of a twist angle between graphene and \nbse, and of spin-orbit coupling and superconducting pairing in \nbse.
We obtain the parameters entering the continuous model via ab-initio calculations.
We show that despite the large mismatch between the graphene's and \nbse's lattice constants,
due to the large size of the \nbse's Fermi pockets, there is a large range of values of twist angles
for which a superconducting pairing can be induced into the graphene layer.
In addition, we show that the superconducting gap induced into the graphene is extremely robust to
an external in-plane magnetic field. 
Our results show that the size of the induced superconducting gap, and its robustness against in-plane magnetic fields, 
can be significantly tuned by varying the twist angle.
\end{abstract}


\maketitle

\section{Introduction}

%
Transition metal dichalcogenides (TMDs) are extremely interesting materials due to their unique electronic properties~\cite{ashwin2012,latzke2015,guinea2014,strano2012,wangzhe2015,gmitra2016,reed2014,ugeda2015,xi2016,xiaoxiang2015,xiaodong2014}
and the fact that in recent years experimentalists have been able to isolate and probe TMD films only few atoms thick, down to the monolayer limit.
Some TMDs monolayers, like \mose and \mos, are insulators with gaps of the order of 1.5-2~eV.
Other TMDs monolayers, such as \nbse, \nbs, \tase, \tas are metallic at room temperature and superconducting at low temperature.
One feature that all TMDs have in common is a strong spin-orbit coupling (SOC).
In monolayer TMDs the strongest effect of the SOC is a spin-splitting of the conduction
and valence bands around the $K$, and $K'$, points of the Brillouin zone (BZ)~\cite{boker2001,ding2011,mak2010}.
For the TMDs that are superconducting at low temperature, such a spin splitting causes 
the superconducting pairing to be of the Ising type~\cite{xi2016} and therefore extremely robust to
external in-plane magnetic fields~\cite{steinberg2018,lu2015,saito2015,barrera2018}.
The ability of metallic TMDs to exhibit superconductivity even in the limit in which they are only one-atom thick,
and the robustness of such superconducting state to external magnetic fields make them very interesting systems
both from a fundamental point of view and for possible applications.

Recent advances in fabrication techniques have made possible the realization of van der Waals (vdW) heterostructures obtained by stacking
crystals that are only few atoms thick~\cite{dean2010,geim2013}
In these structures the different layers are held together by van der Waals forces.
As a consequence the crystals that can be used to create the structures, and their stacking configuration,
are not limited to the configurations allowed by chemical bonds.
%
%
%
This makes possible the realization of systems with unique 
properties
such as graphene--topological-insulator heterostructures
in which graphene has a tunable spin-orbit coupling depending on the stacking configuration~\cite{liangzhi2013,liang2016,junhua2014,zalic2017,rodriguez2017}.
%

In graphene the conduction and valence bands touch at the corners ($K$ and $K'$ points) of the hexagonal BZ,
and around these points the electrons behave as massless Dirac Fermions~\cite{novoselov2005,castro2009}.
This fact makes graphene an ideal semimetal in which the polarity of the carriers can easily be tuned
via external gates. In addition, graphene has a very high electron mobility due to its very low concentration of defects
and the fact that electron-phonons scattering processes do not contribute significantly to the resistivity
for temperatures as high as room temperature~\cite{rossi2009,rossi2011,rodriguez2014}.
%
%
All these features make graphene an ideal system to probe, via tunneling setups, other materials 
and to realize novel vdW heterostructures with tunable properties.
In particular, the fact, that the low energy states of graphene, in momentum space, are located just at the $K$ points of its BZ, 
in vdW structures implies that by simply varying the twist angle, graphene can be used as a momentum selective
probe of the electronic structure, and properties, of the substrate. The work that we present below is
an example of such momentum-selective probing capability of graphene.
%
%
In monolayer \nbse the Fermi surface (FS) is formed by a pocket around the $\Gamma$ point,
and pockets around the $K$, and $K'$ points. Contrary to bulk \nbse, in monolayer \nbse
there is no selenium-like FS pocket around the $\Gamma$ point. As a consequence
monolayer \nbse is expected to be a single-gap superconductor with the same gap at the $\Gamma$ pocket 
as at the $K$ pockets~\cite{Khestanova2018}.
However, the $\Gamma$ and $K$ pockets differ in the magnitude, and $k$ dependence around the pocket,
of the spin-splitting induced by the spin-orbit coupling. The splitting is much larger for the $K$ pockets
and therefore the superconducting gap for these pockets is much more robust to
external in-plane magnetic fields than for the $\Gamma$ pocket.
As we show below a graphene-\nbse heterostructure allows to probe separately \nbse's states around the $\Gamma$ point, and $\KK$ point
simply by tuning the relative twist angle between graphene and \nbse and therefore to study the difference between pockets of the
interplay between spin-orbit coupling and superconducting pairing.

In this work we study vdW heterostructures formed by graphene and monolayer \nbse. Our results show that despite the large 
mismatch between the lattice constants of graphene and \nbse in these structures a large superconducting pairing
can be induced into the graphene layer. In addition, we show how such pairing depends, both in nature and structure,
on the stacking configuration. Our results are relevant also to other graphene-TMD heterostructures such as the ones
that can be obtained by replacing the \nbse monolayer by a monolayer of \nbs, \tase, or \tas that have also been shown
to be superconducting at low temperature~\cite{leroux2012,samuely2010,barrera2018,navarro2016} and show
how graphene can be used to probe in these systems the momentum-dependent  superconducting gap and in particular its
multiband structure.

\section{Method}
\begin{figure}[!h]
 \begin{center}
  \includegraphics[width=\columnwidth]{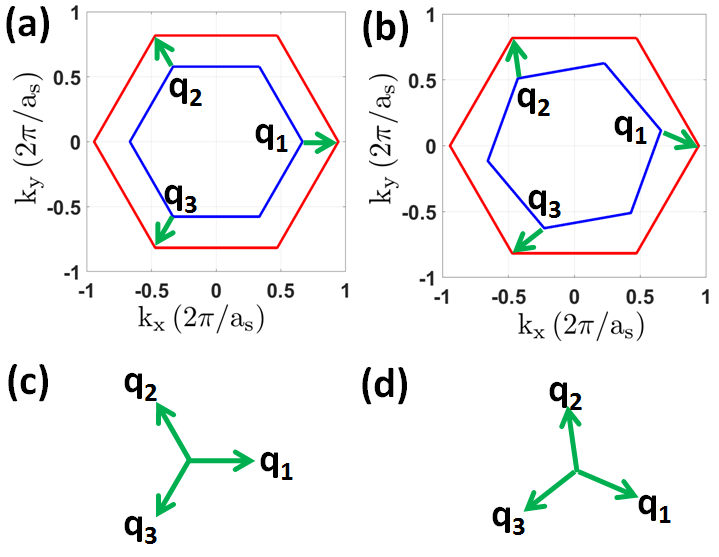}
  \caption{Brillouin zone for graphene and \nbse, and corresponding q-vectors for the case 
          when $\theta=0$, (a), (c), and $\theta\neq 0$, (b), (d).
         } 
  \label{fig.bz}
 \end{center}
\end{figure}

%
In graphene the carbon atoms are arranged in a 2D hexagonal structure formed by two triangular sublattices, $A$ and $B$, with lattice
constant $a_g=\sqrt{3}a=2.46\ang$, with $a=1.42\ang$ the carbon-carbon atomic distance.
The 2D structure of \nbse is also formed by two triangular sublattices. One of the sublattices is formed by the Nb atoms,
the other by two Se atoms symmetrically displaced by a distance $u=1.679\ang$ above and below the plane formed by the Nb atoms.
The lattice constant of \nbse is $a_s=3.48\ang$.~\cite{ding2011}.
Figure~\ref{fig.bz} shows the Brillouin zone of graphene and \nbse. In this figure and in the remainder we take 
$k_x$ to be in the direction connecting the valley $\KK$ with its time-reversed partner $\KK'$.
Figure~\ref{fig.bz}~(a) shows the relative orientation of the graphene's and \nbse's BZs for the case when the twist angle $\theta$
is zero and Figure~\ref{fig.bz}~(b) for a case when $\theta\neq 0$.

To obtain the electronic structure of the graphene-\nbse structure for a generic twist angle and in the presence of superconducting
pairing in the \nbse, we first need to estimate the charge transfer between the graphene layer and \nbse, and the strength
of the tunneling $t$ between graphene and the \nbse monolayer.
To this effect we first obtain via ab-initio the electronic structure of a commensurate graphene-\nbse structure.
Let
$\aav_{1s}=a_s[\cos(\pi/3-\theta)\xx-\sin(\pi/3-\theta)\yy]$,
$\aav_{2s}=a_s[\cos(\pi/3 + \theta)\xx+\sin(\pi/3 + \theta)\yy]$,
be the primitive lattice vectors for \nbse,
and 
$\aav_{1g}=a_g[\cos(\pi/3)\xx-\sin(\pi/3)\yy]$,
$\aav_{2g}=a_g[\cos(\pi/3)\xx+\sin(\pi/3)\yy]$,
the primitive vectors for graphene,
with $\xx$ and $\yy$ the unit vectors in the $x$ and $y$ direction, respectively.
In a commensurate stacking configuration the primitive vectors satisfy the equation:
\beq
 m_1 \aav_{1s} +  m_2 \aav_{2s} = n_1 \aav_{1g} + n_2 \aav_{2g}
 \label{eq:comm01}
\enq
where $(m_1, m_2, n_1, n_2)$ are four integers constrained by the following second order Diophantine equation:
\beq
 (m_1^2 + m_2^2 - m_1 m_2) = \frac{a_g^2}{a_s^2}(n_1^2 + n_2^2 -n_1 n_2).
 \label{eq:comm02}
\enq 
Given that the lattice constant of graphene and NbSe are highly incommensurate with respect to each other, Eq.~\ref{eq:comm01} (or, equivalently, Eq.~\ref{eq:comm02}) can only be satisfied for structures with primitive cells comprising a very large number of atoms ($\sim$1000). It is computationally extremely expensive to study structures with such large primitive cells using ab-initio methods. For this reason we allow for few percents strain of the graphene's lattice so that Eq.~\ref{eq:comm01} (or, equivalently, Eq.~\ref{eq:comm02}) can be satisfied for structures with primitive cells comprising 100 atoms or less.  In general the relative strain of the graphene's and NbSe2's lattices will depend on the specific structure considered. We did not perform an energy minimization analysis and chose to strain graphene rather than NbSe2 for convenience. This is justified considering that the amount of charge transfer between the graphene layer and NbSe2, and the magnitude of the graphene-NbSe2 tunneling strength, are not expected to be affected by a small change of the graphene’s or NbSe2's lattice constant

%
%

The ab-initio calculation were performed using the Quantum-Espresso package~\cite{QE-2009,QE-2017}.
We use full-relativistic ultrasoft pseudopotentials with the wavefunction kinetic energy cutoff of 50~Ry. We adopted the Perdew-Burke-Ernzerhof (PBE) \cite{Perdew1996} as the exchange and correlation functional. 
We set the vacuum thickness equal to $25 \mathrm{\AA}$ to isolate the heterostructure and avoid the interactions between the periodic layers along the direction, ($z$), perpendicular to the layers.
The interlayer distance between graphene and \nbse was obtained by full relaxation in the z-direction.
The total energy was calculated by using a $18\times18\times1$ Monkhorst-Pack scheme grid for the $k$ points. 

After having obtained the amount of charge transfer and the strength of the tunneling between the graphene layer and
\nbse via ab-initio, we use a continuum model 
\cite{mele2010,mele2012,bistritzer2011,junhua2014}
to obtain the low-energy spectrum of the graphene-\nbse heterostructure for different values of the twist angle $\theta$.
%
%
In general, the Hamiltonian $\hat H$ describing the graphene-\nbse heterostructure can be written as:
$\hat H = \hat H_g + \hat H_s + \hat H_t$
where $\hat H_g$ is the Hamiltonian for graphene, 
$\hat H_s$ is the Hamiltonian for \nbse and $\hat H_t$ is the term describing tunneling processes
between graphene and \nbse.
%

In graphene the low energy states are located at the $\KK_g$ and $\KK'_g$ points of the BZ:
$\KK_g=(4\pi/(3a_g),0)$, $\KK'_g=(-4\pi/(3a_g),0)$ (and equivalent points connected by reciprocal lattice wave vectors).
Close the $\KK_g$ and $\KK'_g$ points in graphene the electrons, at low energies, are well described as massless Dirac fermions
with Hamiltonians 
$\hat H_{\KK_g} = \sum\limits_{\kk,\tau\tau'\sigma\sigma'}c^\dagger_{\KK_g+\kk,\tau'\sigma'}  H_{\KK_g}  c_{\KK_g+\kk,\tau\sigma}$, 
$\hat H_{\KK'_g}= \sum\limits_{\kk,\tau\tau'\sigma\sigma'}c^\dagger_{\KK'_g+\kk,\tau'\sigma'} H_{\KK'_g} c_{\KK'_g+\kk,\tau\sigma}$,
where 
\begin{align}
 H_{\KK_g} &= \hbar v_F\kk\cdot\stau\sigma_0-\mu_g\tau_0\sigma_0,
 \label{eq:HgK} \\
 H_{\KK'_g}&=-\hbar v_F\kk\cdot\stau^*\sigma_0-\mu_g\tau_0\sigma_0,
 \label{eq:HgKp} 
\end{align}
$c^\dagger_{\pp,\tau\sigma}$ ($c_{\pp,\tau\sigma}$) is the creation (annihilation) operator for an electron, in the graphene sheet,
with spin $\sigma$ and two-dimensional momentum $\hbar\pp=\hbar(p_x,p_y)$,
$\kk$ is a wave vector measured from $\KK$ ($\KK'$),
$v_F=10^6$~m/s is graphene's Fermi velocity, $\mu_g$ graphene's chemical potential,
and ${\tau_i}$, ${\sigma_i}$ ($i=0,1,2,3)$) are the $2\times2$ Pauli matrices in sublattice and spin space, respectively.
As a consequence, when considering the states of graphene close to the $\KK_g$ ($\KK'_g$) point 
we have $H_g=H_{\KK_g}$ ($H_g=H_{\KK'_g}$).

In \nbse the low energy states are located close to the $\Gamma$, $\KK$, and $\KK'$ points of the BZ:
$\KK_s=(4\pi/(3a_s),0)$, $\KK'_s=(-4\pi/(3a_s),0)$ (and equivalent points connected by reciprocal lattice wave vectors).
%
Close the $\Gamma$ point the effective low-energy Hamiltonian for \nbse takes the form
$H_{\Gamma_s} = \sum_{\kk\sigma\sigma'}d^\dagger_{\kk,\sigma} H_{\Gamma_s}  d_{\kk,\sigma'}$,
%
where $d^\dagger_{\kk,\sigma}$ ($d_{\kk,\sigma}$) is the creation (annihilation) operator for an electron in \nbse 
with momentum $\kk$ and spin $\sigma$, and $H_{\Gamma_s}$ is the effective low energy Hamiltonian matrix
for the conduction band of \nbse.
By fitting the ab-initio results we obtain:
\beq
 H_{\Gamma_s} = \epsilon_{0\Gamma}(\kk) \sigma_0 + \lambda_\Gamma(\kk) \sigma_z
 \label{eq:HnG}
\enq
where
%
%
\begin{align}
 \epsilon_{0\Gamma}(\kk) &= \eta_{0\Gamma} + \eta_{2\Gamma} k_+ k_- \nonumber \\
 \lambda_\Gamma(\kk) &= \l_{3\Gamma} \left[ \left(k_+^3 + k_-^3 \right)\cos(3\theta) + i \left(k_+^3 - k_-^3 \right)\sin(3\theta) \right],
\label{eq:HnbseG}
\end{align}
$k_{\pm}=k_x\pm i k_y$, and $\eta_{0\Gamma}$, $\eta_{2\Gamma}$,  $\l_{3\Gamma}$ are constants:
\begin{align}
 \eta_{0\Gamma} &= \phantom{-}0.5641\:  \mathrm{eV}, \nonumber \\
 \eta_{2\Gamma} &= -7.0640\: \mathrm{eV} \:[a_s/(2\pi)]^2,  \nonumber \\
 l_{3\Gamma}    &=\phantom{-}0.5085 \: \mathrm{eV} \: [a_s/(2\pi)]^3. 
\label{eq:HnbseGpars}
\end{align}

Close to the corners of the BZ of \nbse, the $\KK_s$ and $\KK'_s$ points,
for \nbse we have
$H_{\KK_s} = \sum_{\kk\sigma\sigma'}d^\dagger_{\kk,\sigma} H_{\KK_s}  d_{\kk,\sigma'}$,
$H_{\KK'_s} = \sum_{\kk\sigma\sigma'}d^\dagger_{\kk,\sigma} H_{\KK'_s}  d_{\kk,\sigma'}$,
where $\kk$ is now a wave vector measured from the $\KK_s$, $\KK'_s$ point, respectively, and
\begin{align}
 H_{\KK_s} &= \epsilon_{0}(\kk) \sigma_0 + \epsilon_3(\kk)\sigma_0 + \lambda(\kk) \sigma_z
 \label{eq:HnK} \\
 H_{\KK'_s} &=\epsilon_{0}(\kk) \sigma_0 - \epsilon_3(\kk)\sigma_0 - \lambda(\kk) \sigma_z 
 \label{eq:HnKp} \\
\end{align}
where,
\begin{align}
 \epsilon_{0}(\kk)  &= \eta_{0} + \eta_{2} k_+ k_-, \nonumber \\
 \epsilon_3(\kk)         &=\eta_3\left[ \left(k_+^3 + k_-^3 \right)\cos(3\theta) + i \left(k_+^3 - k_-^3 \right)\sin(3\theta) \right],\nonumber \\
 \lambda(\kk) &=l_0+ \l_{2} k_+ k_-,
\label{eq:HnbseK}
\end{align}
and $\eta_0$, $\eta_2$, $\eta_3$, $l_0$, $l_2$, are constants that we extracted from the ab-initio results for an isolated monolayer of \nbse: 
\begin{align}
 \eta_{0} &= \phantom{-}0.4526\:  \mathrm{eV}, \nonumber \\
 \eta_{2} &= -9.0940\: \mathrm{eV} \:[a_s/(2\pi)]^2,  \nonumber \\
 \eta_{3} &= \phantom{-}3.07\phantom{00}\: \mathrm{eV} \:[a_s/(2\pi)]^3,  \nonumber \\
 l_{0}    &=\phantom{-}0.0707 \: \mathrm{eV},\nonumber \\
 l_{2}    &=-0.33\phantom{00} \: \mathrm{eV} \: [a_s/(2\pi)]^2. 
\label{eq:HnbseKpars}
\end{align}
%


Let $\pp_g$, $\pp_s$, be the wave vector of an electron in graphene, \nbse, respectively.
In the remainder we consider only momentum and spin conserving tunneling processes.
Conservation of crystal momentum requires
\beq
 \pp_s + \GG_s = \pp_g + \GG_g, 
 \label{eq:mom_con01}
\enq
where $\GG_g$ and $\GG_s$ are reciprocal lattice vectors for graphene and \nbse respectively.
%
For the purpose of developing a continuum low energy model for a graphene-\nbse heterostructure
it is more convenient to consider the twist angle $\theta$ as relative twist between BZ's,
as shown in Fig.~\ref{fig.bz}. For $\theta=0$ the $\KK$ point of graphene's and \nbse's BZs
are on the same axis.
Depending on the value of $\theta$ we can have two situations:
the low energy states of graphene, in momentum space, are close to \nbse's Fermi pockets around the $\KK$ and $\KK'$ points,
or, considering \nbse's extended BZ, to \nbse's Fermi pocket around the $\Gamma$ point.
In the first case the conservation of the crystal momentum, Eq.~\ceq{eq:mom_con01}, takes the form:
\beq
 \kk_s = \kk_g + (\KK_g-\KK_s) + (\GG_g - \GG_s)
 \label{eq:mom_conK}
\enq
where $\kk_s$ $\kk_g$ are momentum wave vectors measured from $\KK_g$ and $\KK_s$, respectively.
By replacing $\KK_g$, $\KK_s$, with $\KK'_g$, and $\KK'_s$ in Eq.~\ceq{eq:mom_conK}
we obtain the momentum conservation equation valid for momenta taken around the $\KK'$ points.
In the second case Eq.~\ceq{eq:mom_con01} takes the form:
\beq
 \kk_s = \kk_g + \KK_g + (\GG_g - \GG_s)
 \label{eq:mom_conG}
\enq
and similarly for momenta around $\KK'_g$.

The conservation of the crystal momentum implies 
that the tunneling term takes the form:
\beq
 \hat H_t = \!\!\!\!\!\!\!\!\sum_{\GG_g\GG_s\\ \tau\sigma}\!\!\!\! \hat T_{\tau\sigma\sigma'}(\pp_g + \GG_g)e^{-i\GG_g\cdot\dd_\tau}c^\dagger_{\pp_g\tau\sigma}
           d_{\pp_g + (\GG_g - \GG_s)\sigma'} + h.c.
 \label{eq:H_t}
\enq
where $\dd_\tau$ is the position of the carbon atom on sublattice $\tau$ within the primitive cell of the graphene sheet.
For sublattice  $A$ $\dd_\tau=(0,0)$, for sublattice $B$ $\dd_\tau=(a_0,0)$, with $a_0$ the carbon-carbon distance.

Considering that, as shown in table~\ref{tab:commensurate}, the separation $d=3.57 \AA$ between the graphene sheet and \nbse is much larger
than the interatomic distance in each material, in momentum space, the 
tunneling amplitude $t(\pp)$ decays very rapidly as a function of $\pp$~\cite{bistritzer2011}
and so in Eq.\ceq{eq:H_t}  we can just keep the terms for which $(\pp_g+\GG_g)$ is smallest, i.e., restrict the sum
to $\GG_g=0$ and the two $\GG_g$ that map $\KK$ ($\KK'$) to the two other equivalent points in the BZ and
set $t=t(\KK)$. The sum over $\GG_s$ is restricted by the fact that we only need to keep terms for which 
the graphene's and \nbse's states have energy separated by an amount of the order of $t$. 
%
%

Let $\qq=\kk_s-\kk_g$.
The above considerations imply that
for the case when the $\KK_g$ and $\KK_s$ are close we only need to keep the terms for which 
$|\qq|=|\KK_g-\KK_s|$, given that these are the terms for which $(\pp_g+\GG_g)$ that satisfies Eq.~\ceq{eq:mom_conK} is smallest.
Due to the  $C_{3v}$ symmetry of the hexagonal structure there 
are three equivalent $\KK$ points, $\KK_1$, $\KK_2$, $\KK_3$, (and $\KK'$ points),
i.e. two reciprocal lattice wave vectors $\GG$ connecting equivalent corners of the BZ.
There are three vectors 
$\qq_{iK}= (\KK_g-\KK_s) + (\GG_{gi} - \GG_{si})$ ($i=1,2,3$) such that $|\qq_i|=|\KK_g-\KK_s|$.
$\qq_{1K}$ is obtained by taking $\GG_{g1}=0$ and 
$\GG_s=\GG_{sK1}\df 0$, 
$\qq_{2K}$ by taking 
$\GG_g=\GG_{g2}\df 4\pi/(\sqrt{3}a_g)[\cos(5\pi/6),\sin(5\pi/6)]$,  
$\GG_s=\GG_{sK2}\df 4\pi/(\sqrt{3}a_s)[\cos(5\pi/6+\theta),\sin(5\pi/6+\theta)]$,
and
$\qq_{3K}$ by taking 
$\GG_g=\GG_{g3}\df 4\pi/(\sqrt{3}a_g)[\cos(7\pi/6),\sin(7\pi/6)]$,  
$\GG_s=\GG_{sK3}\df 4\pi/(\sqrt{3}a_s)[\cos(7\pi/6+\theta),\sin(7\pi/6+\theta)]$.

When the graphene's low energy states are close to the $\Gamma$ pocket of \nbse's second BZ
the smallest possible value of $|\qq|$ is $|\KK_g-\GG_s|$ with
$\GG_s=4\pi/(\sqrt{3}a_s)[\cos(-\pi/6+\theta),\sin(-\pi/6+\theta)]$.
As before, considering the $C_{3v}$ symmetry,
there are three vectors $\qq_{i\Gamma}$ with this magnitude:
$\qq_{1\Gamma}$ obtained by 
taking 
$\GG_g=0$, 
$\GG_s=\GG_{s\Gamma1}\df 4\pi/(\sqrt{3}a_s)[\cos(-\pi/6+\theta),\sin(-\pi/6+\theta)]$,
$\qq_{2\Gamma}$ obtained by 
taking 
$\GG_g=\GG_{g2}$,
$\GG_s=\GG_{s\Gamma2}\df 4\pi/(\sqrt{3}a_s)[\cos(\pi/2+\theta),\sin(\pi/2+\theta)]$,
and 
$\qq_{3\Gamma}$ obtained by 
taking 
$\GG_g=\GG_{g3}$,
$\GG_s=\GG_{s\Gamma3}\df 4\pi/(\sqrt{3}a_s)[\cos(7\pi/6+\theta),\sin(7\pi/6+\theta)]$,

By retaining only the tunneling terms for which $t(\pp_g + \GG_g)$ is largest, when considering
the graphene states close to the $\KK_g$ point so that $H_g=H_{\KK_g}$, 
we can rewrite $\hat H_t$ as
%
%
\beq
 \hat H_t = \sum_{i=1}^3 c^\dagger_{\kk_g\tau\sigma}T^\dagger_{\KK_g,i,\tau\sigma\sigma'} d_{\kk_g + \qq_i,\sigma'} + h.c.     
 \label{eq:Ti}
\enq
with:
\begin{eqnarray}
T_{\KK_g,1}^\dagger = \left[
\begin{array}{cccc}
t     &     0      &      t        &     0   \\
0     &     t      &      0        &     t
\end{array}
\right]
\label{Eq.hopping_T1}
\end{eqnarray}
\begin{eqnarray}
T_{\KK_g,2}^\dagger = \left[
\begin{array}{cccc}
t     &     0      &      t e^{-i\GG_{g2}\cdot\dd_B}        &     0   \\
0     &     t      &      0        &     t e^{-i\GG_{g2}\cdot\dd_B}
\end{array}
\right]
\label{Eq.hopping_T2}
\end{eqnarray}
\begin{eqnarray}
T_{\KK_g,3}^\dagger = \left[
\begin{array}{cccc}
t     &     0      &      t e^{-i\GG_{g3}\cdot\dd_B}        &     0   \\
0     &     t      &      0        &     t e^{-i\GG_{g3}\cdot\dd_B}
\end{array}
\right].
\label{Eq.hopping_T3}
\end{eqnarray}
%

In the remainder, supported by DFT results, we take $t$ to be the same both
when the graphene's low energy states tunnel into states around the $\KK$ ($\KK'$) point and the $\Gamma$ point of \nbse.
Let $\gamma\df t/ \hbar v_F|\qq_i|$. 
When $\gamma< 1$ we can develop a perturbative approach in which $\gamma$ is the small parameter~\cite{dossantos2007,bistritzer2011}:
terms of order $\gamma^n$ correspond n-tuple tunneling processes.
For our situation, as we show in the following section, $\gamma\ll 1$ and so we can retain just the lowest order terms in $\gamma$.

It is convenient to define the following spinors:
\begin{align}
  C^\dagger_\kk &= (c^\dagger_{\kk A\uparrow}, c^\dagger_{\kk A\downarrow}, c^\dagger_{\kk B\uparrow},c^\dagger_{\kk B\downarrow}); \nonumber \\
  D^\dagger_{\Gamma\kk} &= (d^\dagger_{\kk \uparrow}, d^\dagger_{\kk \downarrow}); \nonumber \\
  D^\dagger_{K,\kk} &= (d^\dagger_{\KK_s+\kk \uparrow}, d^\dagger_{\KK_s+\kk \downarrow}); 
  \nonumber \\
  \Psi^\dagger_{K_g\Gamma_s\kk} &= (C^\dagger_\kk,  D^\dagger_{\Gamma,\kk+\qq_{1\Gamma}},  D^\dagger_{\Gamma,\kk+\qq_{2\Gamma}},  D^\dagger_{\Gamma,\kk+\qq_{3\Gamma}}); \nonumber \\
  \Psi^\dagger_{K_gK_s,\kk} &= (C^\dagger_\kk,  D^\dagger_{K,\kk+\qq_{1K}},  D^\dagger_{K,\kk+\qq_{2K}},  D^\dagger_{K,\kk+\qq_{3K}}).
  \nonumber
\end{align}
For the case when the graphene's FS overlaps with the \nbse's pocket close to the $K$ point,
we can then express the Hamiltonian for the graphene-\nbse system as 
$\hat H_{K_gK_s}=\sum_\kk \Psi^\dagger_{\kk,K_gK_s} H_{K_gK_s}(\kk) \Psi_{\kk,K_gK_s}$
%
%
%
with
\begin{widetext}

\begin{eqnarray}
H_{K_gK_s}(\kk) = \left[
\begin{array}{cccc}
H_{\KK_g}(\kk)      &  T_{\KK_g,1}               &   T_{\KK_g,2}                 &   T_{\KK_g,3}  \\
T_{\KK_g,1}^\dagger   &  H_{\KK_s+\GG_{sK1}}(\kk+\qq_{1K})      &   0                   &    0    \\
T_{\KK_g,2}^\dagger   &     0               &   H_{\KK_s+\GG_{sK2}}(\kk+\qq_{2K})       &    0    \\
T_{\KK_g,3}^\dagger   &     0               &   0                   & H_{\KK_s+\GG_{sK3}}^S(\kk+\qq_{3K})
\end{array}
\right].
\label{eq:HK}
\end{eqnarray}
%

%

\end{widetext}

For the case when we consider graphene states close to the $\KK'_g$ point, so that $H_g=H_{\KK'_g}$,
the expression of the Hamiltonian matrix $H_{K'_gK'_s}(\kk)$ for the graphene-\nbse system, within the approximations described above, 
can be obtained from Eq.~\ceq{eq:HK} by doing the following substituions:
$\KK_s\to\KK'_s$, $\GG_{gi}\to -\GG_{gi}$, $\GG_{si}\to -\GG_{si}$, $\qq_{iK}\to -\qq_{iK}$ and noticing that
$T_{\KK'_g,i}=T^*_{\KK_g,i}$.
Similarly, when the low energy states of graphene are close to the $\Gamma$ point of \nbse 
the Hamiltonian $H_{K_g\Gamma}(\kk)$ ($H_{K'_g\Gamma}(\kk)$) is obtained from
the expression~\ceq{eq:HK} for $H_{K_gK_s}(\kk)$ via the substitutions
$\KK_s+\GG_{sKi}\to\GG_{s\Gamma i}$ 
($\KK'_s-\GG_{sKi}\to-\GG_{s\Gamma i}$),
and 
$\qq_{iK}\to\qq_{i\Gamma}$
($\qq'_{iK}\to-\qq_{i\Gamma})$.

Including the superconducting pairing, the effective low-energy Hamiltonian for \nbse for states close 
to the $\Gamma$ point takes the form
\beq
 \hat H_{\Gamma_s}^{(\rm SC)} = \sum_{\kk}\Psi^\dagger_{\kk s}  H_{\Gamma_s}^{(\rm SC)}\Psi_{\kk s}, 
\enq
where $\Psi^\dagger_{\kk s}$ is the Nambu spinor
$\Psi^\dagger_{\kk s}=(D^\dagger_{\kk},D_{-\kk})$,
\begin{eqnarray}
 H_{\Gamma_s}^{(\rm SC)}  = \left[
 \begin{array}{cc}
   H_{\Gamma_s}(\kk)          &  i \Delta_\Gamma \sigma_2 \\
 -i\Delta_\Gamma\sigma^*_2   &  -H^T_{\Gamma_s}(-\kk)                      
\end{array}
\right],
\label{eq:HsGSC}
\end{eqnarray}
$H_{\Gamma_s}(\kk)$ is given by Eq.~\ceq{eq:HnG},
and $\Delta_\Gamma$ is the size of the superconducting  gap of \nbse close to the $\Gamma$ point.

For states close to $\KK_s$, including the superconducting pairing, the Hamiltonian for \nbse becomes
\beq
 \hat H_{sK}^{(\rm SC)} = \sum_{\kk_n}\Psi^\dagger_{\kk_s}  H_{sK}^{(\rm SC)}\Psi_{\kk_s}, 
\enq
where now $\kk$ ($-\kk$) is understood to be measured from $\KK_s$ ($\KK'_s$), and
%
%
\begin{eqnarray}
 H_{sK}^{(\rm SC)}  = \left[
 \begin{array}{cc}
   H_{s\KK_s}(\kk)          &  i \Delta_K \sigma_2 \\
 -i\Delta_K\sigma^*_2   &  -H^T_{s\KK'_s}(-\kk)                      
\end{array}
\right],
\label{eq:HsKSC}
\end{eqnarray}
$H_{s\KK_s}(\kk)$, $H_{s\KK'_s}(\kk)$ are given by Eq.~\ceq{eq:HnK}.

For monolayer \nbse the superconducting gap is expected to have the same value, $\Delta$, on the $\Gamma$ and $K$ pocket.
In the remainder we conservatively assume $\Delta=0.5$~meV~\cite{Khestanova2018}.

The Hamiltonian for the graphene-\nbse system including the superconducting pairing 
in \nbse. For the case when
$\KK_g$ is close to $\KK_s$ the 
Hamiltonian becomes $\hat H_{K_gK_s}^{(\rm SC)}= \sum_{\kk}\Psi^\dagger_{K_g K_s, SC,\kk}  H_{K_gK_s}^{(\rm SC)}(\kk)\Psi_{K_g K_s, SC,\kk}$,
with
$\Psi^\dagger_{K_g K_s, SC,\kk} =(\Psi^\dagger_{K_g K_s,\kk}, \Psi^T_{K'_g K'_s,-\kk}$),
%
%
%
%
\begin{eqnarray}
 H_{K_gK_s}^{(\rm SC)}(\kk)  = \left[
 \begin{array}{cc}
   H_{K_gK_s}(\kk)          &  \Delta_K \Lambda \\
   \Delta_K\Lambda^\dagger  &  -H^T_{K'_gK'_s}(-\kk)                      
\end{array}
\right],
\label{eq:HKSC}
\end{eqnarray}
and
\begin{eqnarray}
 \Lambda  = \left[
 \begin{array}{cccc}
   0_{4\times 4} & 0_{4\times 2} & 0_{4\times 2} & 0_{4\times 2} \\
   0_{2\times 4} & i\sigma_2  & 0_{2\times 2} & 0_{2\times 2} \\
   0_{2\times 4} & 0_{2\times 2} & i\sigma_2 & 0_{2\times 2} \\
   0_{2\times 4} & 0_{2\times 2} & 0_{2\times 2} & i\sigma_2 \\       
\end{array}
\right]
\label{eq:lambda}
\end{eqnarray}
where $0_{m\times n}$ is the zero matrix with $m$ rows and $n$ columns.

Similarly, for the case when the low energy states of graphene are close to the $\Gamma$ point of the extended BZ of \nbse the Hamiltonian for the 
whole system becomes
 $\hat H_{K_g\Gamma_s}^{(\rm SC)}= \sum_{\kk}\Psi^\dagger_{K_g \Gamma_s, SC,\kk}  H_{K_g\Gamma_s}^{(\rm SC)}(\kk)\Psi_{K_g \Gamma_s, SC,\kk}$,
with
$\Psi^\dagger_{K_g \Gamma_s, SC,\kk} =(\Psi^\dagger_{K_g \Gamma_s,\kk}, \Psi^T_{K'_g \Gamma_s,-\kk}$),
%
%
%
%
\begin{eqnarray}
 H_{K_g\Gamma}^{(\rm SC)}(\kk)  = \left[
 \begin{array}{cc}
   H_{K_g\Gamma}(\kk)          &  \Delta_\Gamma \Lambda \\
   \Delta_\Gamma\Lambda^\dagger  &  -H^T_{K'_g\Gamma}(-\kk)                      
\end{array}
\right].
\label{eq:HGSC}
\end{eqnarray}
%


\section{Results}

\begin{figure}[!h]
 \begin{center}
  \centering 
  \includegraphics[width=\columnwidth]{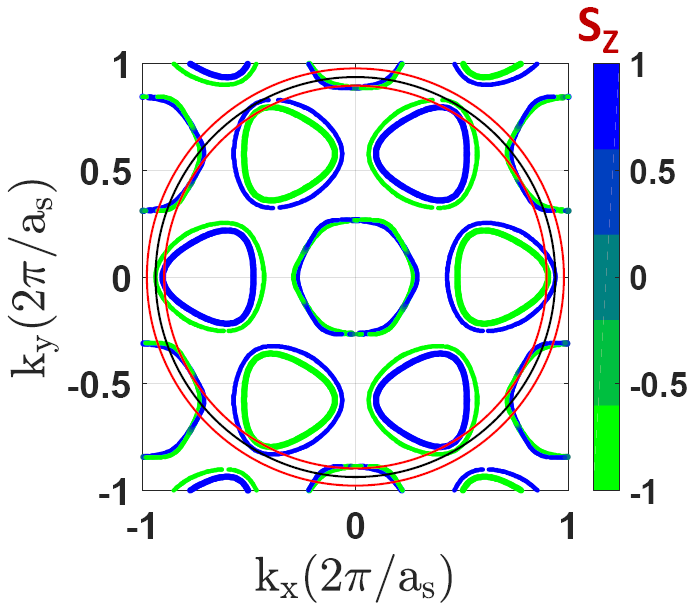}
  \caption{Overlap of the Fermi surfaces of monolayer \nbse and graphene. The blue (green) FSs are the \nbse FSs for spin up (down) respectively,
           the black circle shows the position of the graphene Dirac point for all the possible twist angles,
           and the red circles show the region within which the graphene FS is confined as the twist angle is varied.
         } 
  \label{fig:FS}
 \end{center}
\end{figure}

The large lattice mismatch between graphene and \nbse would suggest that even in the absence of any twist angle the electronic
states of the two systems would not hybridize. However, this does not take into account the large size of \nbse's
Fermi pockets. As shown in Fig.~\ref{fig:FS} there is a large set of values of $\theta$ for which the Dirac point
of graphene intersects the \nbse's FS either around the $K$ points, or around the $\Gamma$ point in the repeated zone scheme.
For these points the electronic states of graphene and \nbse are expected to hybridize.

\begin{figure}[!h]
 \begin{center}
  \centering 
  \includegraphics[width=0.95\columnwidth]{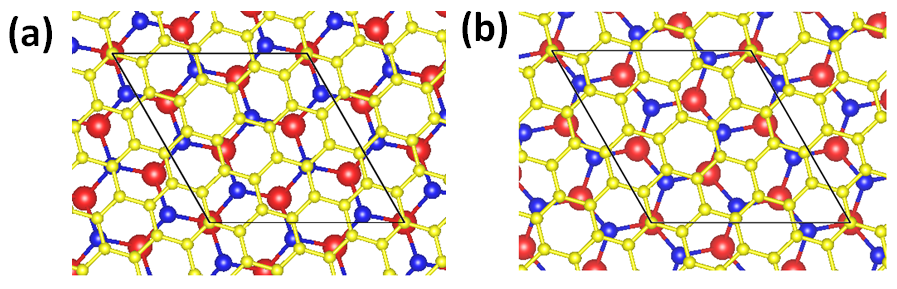}
  \caption{Commensurate graphene-\nbse structure corresponding to the parameters listed in 
           Table~\ref{tab:commensurate}. (a) is the configuration for $\theta = -65.2^0$. (b) is the configuration for $\theta = 33.0^0$.
           The red (blue) spheres show Nb (Se) atoms, the graphene lattice is shown in yellow.
         } 
  \label{fig:commensurate}
 \end{center}
\end{figure}

From the results shown in Fig.~\ref{fig:FS} we see that for small values of $\theta$, we can expect that the graphene's low energy states
close to the Dirac point will hybridize with the \nbse's states close to the $K$ point. 
For values of $\theta$ close to $30\degree$ we see that graphene's states will hybridize with \nbse's states close
to the $\Gamma$ point.
For this reason, to estimate the charge transfer and the strength of the graphene-\nbse tunneling in the two situations,
we performed ab-initio calculations for a commensurate heterostructure with $\theta=-65.2\degree$, and one with $\theta=33.0\degree$.
The parameters identifying these commensurate structures are given in table~\ref{tab:commensurate}
and the corresponding primitive cells are shown in
Fig.~\ref{fig:commensurate}.
%
%
%
\begin{table}[!h]
\centering
\resizebox{\columnwidth}{!}{
\begin{tabular}{c c c c c c c c c}
\hline \hline
TMD & ($m_1, m_2,n_1, n_2$) & $a_s(\mathrm{\AA})$ & $a_g(\mathrm{\AA})$ & $\%\delta a_g$ &   $\theta$  &  $|A|(\mathrm{\AA})$ & $d(\mathrm{\AA})$ & $\mu_G (eV)$ \\ 
\hline
\nbse  & $(-2,1,-4,-3)$  &  3.48~\cite{ding2011}  &  2.55  & $3.7 \%$& $-65.2^0$     &  9.2 &  3.57 & -0.40 \\
\hline
\nbse  & $(-1,2,1,4)$  &  3.48~\cite{ding2011}  &  2.55  & $3.7 \%$ &$33.0^0$     &  9.2 &  3.57 & -0.40  \\
\hline
\end{tabular}}
\caption{Parameters for graphene-\nbse commensurate structures.}
\label{tab:commensurate}
\end{table}
%

\begin{figure}[!h]
 \begin{center}
  \centering 
  \includegraphics[width=0.8\columnwidth]{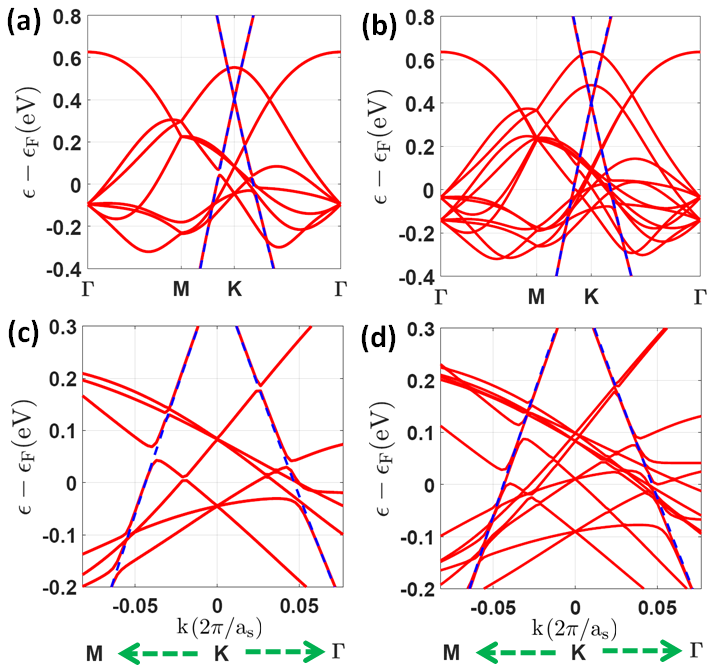}
  \caption{Bands for the commensurate graphene-\nbse structure shown in Fig.~\ref{fig:commensurate}~(a) for which
           $\theta=-65.2\degree$ so that  graphene's FS overlaps with \nbse's FS pocket around the $\KK$ point.
	   (a) No SOC, (b) with SOC. (c): low energy detail of (a). (d): low energy detail of (b).
         } 
  \label{fig:bands01}
 \end{center}
\end{figure}
\begin{figure}[!h]
	\begin{center}
		\centering 
		\includegraphics[width=0.8\columnwidth]{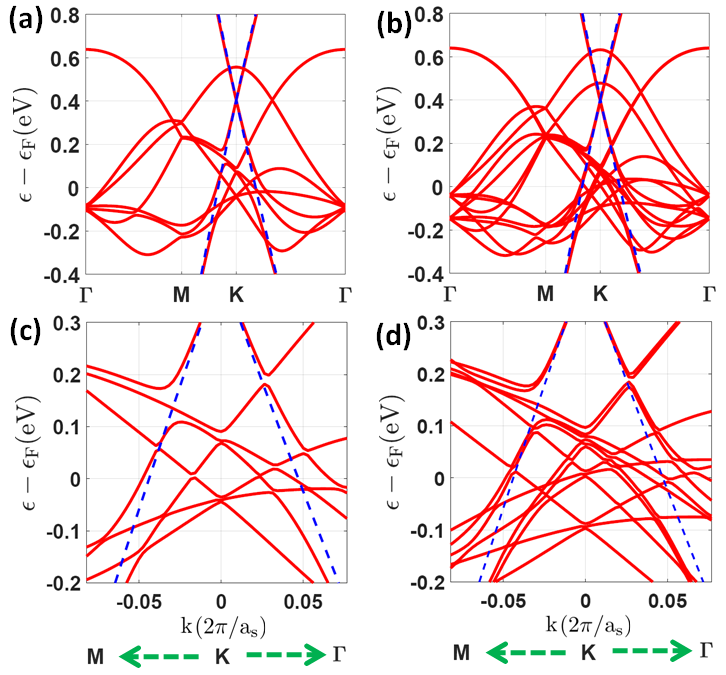}
		\caption{Bands for the commensurate graphene-\nbse structure shown in Fig.~\ref{fig:commensurate}~(b) for which
                  $\theta=33\degree$ so that  graphene's FS overlaps with \nbse's FS pocket around the $\Gamma$ point.
                  (a) No SOC, (b) with SOC. (c): low energy detail of (a). (d): low energy detail of (b).
		} 
		\label{fig:bands02}
	\end{center}
\end{figure}

The ab-initio calculations return the band structure shown in Fig.~\ref{fig:bands01},~\ref{fig:bands02}. 
In these figures the dashed blue lines show the 
bands of isolated graphene. The left panels show the results obtained without including spin-orbit effects and the right panels
the results obtained taking into account the presence of spin orbit coupling.
Panels (c) and (d) show an enlargement at low energies of the results shown in panels (a) and (b).


The results of  Fig.~\ref{fig:bands01},~\ref{fig:bands02} clearly show 
that there is a significant charge transfer between graphene and monolayer \nbse
resulting in hole doping of the graphene sheet corresponding to a Fermi energy of about -0.4~eV.
They also show that the amount of charge transfer does not depend on the value of the twist angle $\theta$.
Considering the finite extension of the graphene's FS due to the charge-transfer shown in Fig.~\ref{fig:bands01}~\ref{fig:bands02}
between \nbse and graphene,
we obtain that there is a significant range of values of $\theta$ for which the graphene's FS intersects the \nbse FS and for which
we can then expect non-negligible hybridization of the graphene's and \nbse states.
This is shown in 
Fig.~\ref{fig:FS} in which the red circles delimit the boundaries of the graphene's FS as $\theta$ is varied.
Table~\ref{table.overlap} shows the range of values of $\theta$ 
extracted from Fig.~\ref{fig:FS}
for which the graphene's FS is expected to intersect
either one of the \nbse's FS pockets around the $K$ ($K'$) point, or around the $\Gamma$ point.
In this table $\theta_m(K)$ ($\theta_m(\Gamma)$) is the angle in the middle of the range $2\delta\theta(K)$ ($2\delta\theta(\Gamma)$)
of angles for which the graphene's FS intersects the \nbse's FS.  
\begin{table}[!h]
\centering
\resizebox{\columnwidth}{!}{
\begin{tabular}{c c c c c }
\hline \hline
TMD (1L) & $\mathrm{\theta_{m}(K)}$ & $\mathrm{\delta\theta(K)}$ & $\mathrm{\theta_{m}(\Gamma)}$ & $\mathrm{\delta\theta(\Gamma)}$ \\ 
\hline
\nbse & $0^0 + n * 60^0$ & $7.2^0$ & $21.9^0 + n * 60^0$ & $3.9^0$ \\
      &                  &         & $37.5^0 + n * 60^0$ & $3.9^0$ \\
\hline
\end{tabular}}
\caption{Values of the twist angle $\theta$ for which the graphene's FS overlap with \nbse's FS pocket around the $K$ point 
         or $\Gamma$ point. For $\mathrm{\theta_m(K)-\delta\theta(K)} \le \theta  \le \mathrm{\theta_m(K)+\delta\theta(K)}$,
         $\mathrm{\theta_m(\Gamma)-\delta\theta(\Gamma)} \le \theta  \le \mathrm{\theta_m(\Gamma)+\delta\theta(\Gamma)}$,
         graphene's FS overlaps \nbse's $K$ pocket, $\Gamma$ pocket, respectively.
         $n$ is an integer between 0 and 5.} 
\label{table.overlap} 
\end{table}

The ab-initio results allow us also to estimate the strength of the tunneling between graphene and \nbse.
In Figs.~\ref{fig:bands01}~(c),~(d),~\ref{fig:bands02}~(c),~(d) we can see the avoided crossings close to the Fermi energy between the graphene's and \nbse's bands.
The amplitude of such crossings provides an estimate of the tunneling strength $t$ between the graphene sheet and the monolayer of \nbse.
We find that both for the case when the graphene's FS intersects the \nbse's pocket around the $\KK$ point and when it intersects the
\nbse's FS pocket around the $\Gamma$ point, $t\approx 20$~meV and so in the remainder we set $t=20$~meV.

\begin{figure}[!h]
 \begin{center}
  \centering 
  \includegraphics[width=\columnwidth]{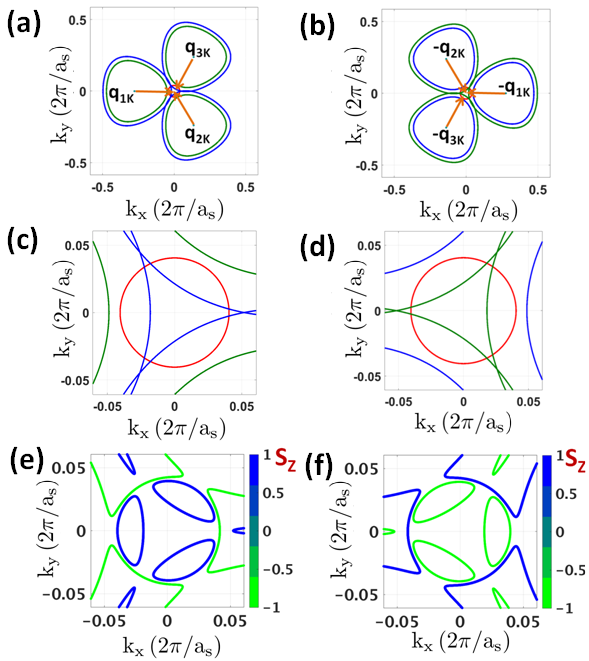}
  \caption{(a) Graphene's FS at the $K$ point (in red) and \nbse's FS (in red and green) for $\theta=0$,
           for which graphene's low energy states are close to \nbse's $K$ point. 
           Because of SOC the \nbse FS for spin-up, shown in blue
           is different from the \nbse's FS for spin down, shown in green.
           The arrows show the vectors $\qq_{iK}$.
           (b) Same as (a) but for graphene's valley around the $K'$point.
           (c), (d) zoom of (a), and (b), respectively.
           (e) FS of graphene-\nbse heterostructure around graphene's $K$ valley 
           for the case when a finite tunneling $t=20$~meV
           between graphene and \nbse is present.
           (f) Same as (e) for graphene's $K'$ valley.
         } 
  \label{fig:K0}
 \end{center}
\end{figure}

We first consider the case when graphene's FS intersects the FS pocket of \nbse close to the $K$ point, i.e. 
$-7.2\degree<\theta<7.2\degree$, and $\Delta=0$.
Figure~\ref{fig:K0} shows the results for the FS of the hybridized system in the limit when no superconducting
pairing is present in \nbse: the left (right) column shows the FS around the $\KK$ ($\KK'$) of graphene.
Figure~\ref{fig:K0}~(a),~(b) show the relative position in momentum space of graphene's FS and \nbse's FS for the case when $\theta=0$ and $t=0$,
taking into account the ``folding'' of the \nbse's FS pockets due to the fact that the three $\KK$ ($\KK'$) corners of the BZ are equivalent.
The graphene FS is shown in red and the spin splitted \nbse's FS in blue and green.
We use this color-convention throughout this work.
A zoom closer to the graphene's $K$ point, Figs~\ref{fig:K0}~(c),~(d), clearly shows the overlap of the graphene's FS with
the \nbse' FS pockets. When $t\neq 0$ the graphene's and \nbse's states hybridize giving
rise to the reconstructed FSs shown in Fig.~\ref{fig:K0}~(e),~(f).
Figures~\ref{fig:K0}~(e),~(f) show that the graphene's FS, due to the hybridization with \nbse,
becomes spin split.

\begin{figure}[!h]
 \begin{center}
  \centering 
  \includegraphics[width=\columnwidth]{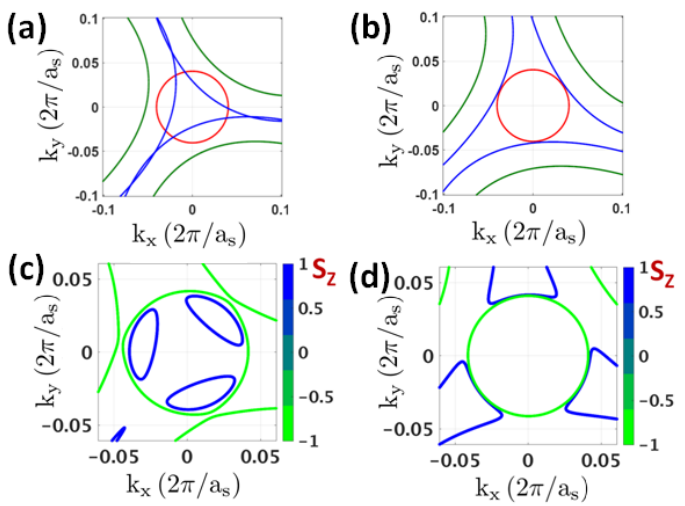}
  \caption{Graphene's and \nbse's FSs for $\theta=2\degree$, (a), and $\theta=6\degree$ in the limit $t=0$.
           (c) FS of graphene-\nbse heterostructure 
           for the case when $t=20$~meV, and $\theta=2\degree$. (d) Same as (c) for $\theta=6\degree$
         } 
  \label{fig:K2K6}
 \end{center}
\end{figure}

Figure~\ref{fig:K2K6} shows the results for the case when $\theta=2\degree$, left column, and $\theta=6\degree$, right columns.
For these values of the twist angle the low energy states of graphene are still close to the low energy states of \nbse located
around \nbse's $K$ points.
For $\theta=2\degree$ the graphene's and \nbse's low energy states are still close enough (in momentum and energy)
that, for $t=20$~meV, the hybridization is strong enough to significantly modify the FS
of the combined system, as shown in Fig~\ref{fig:K2K6}~(c), obtained setting $\Delta=0$.
%
For $\theta=6\degree$ the graphene's and \nbse's FSs are tangent at isolated points
as shown in Fig.~\ref{fig:K2K6}~(b). 
As a consequence, when $t\neq 0$ the states at the FS of graphene and \nbse only hybridize
around these ``tangent-points'', as shown in Fig.~\ref{fig:K2K6}~(d) obtained for $t=20$~meV and $\Delta=0$.


%

\begin{figure}[!h]
	\begin{center}
		\centering 
		\includegraphics[width=\columnwidth]{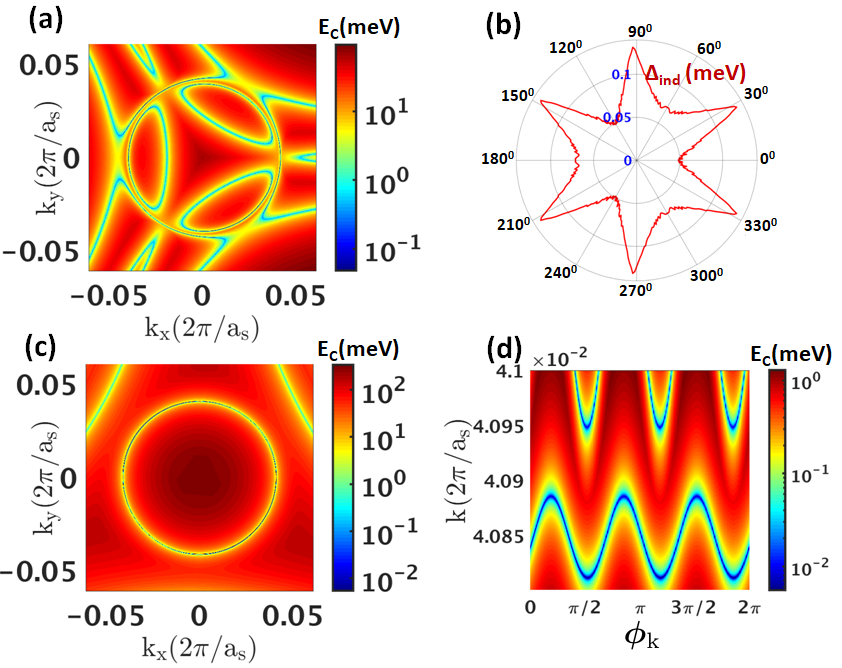}
		\caption{
                         (a) $E_c(\kk)$ for $\theta=0$.
                         (b) $\Delta_{\rm ind}(\phi_k)$ for $\theta=0$.
                         (c) $E_c(\kk)$ for $\theta=9\degree$.
                         (d) $E_c(\phi_k,|\kk|)$ for $\theta=9\degree$ and $|\kk|$ close to the original graphene's Fermi wave vector $k_{F,g}$.
		} 
		\label{fig:delta0_0}
	\end{center}
\end{figure}

We now consider the case when a superconducting gap is present in \nbse.
We find that for $\theta=0$ the FS is completely gapped but the gap is not uniform.
Figure~\ref{fig:delta0_0}~(a) shows the lowest positive electron energy, $E_c$, as a function of $\kk$.
The smallest value of  $E_c(\kk)$ corresponds to the induced superconducting gap $\Delta_{\rm ind}$.
For $\theta=0$ we find $\Delta_{\rm ind}=0.05$~meV.
By calculating the smallest value of $E_c(\kk)$ for each angle $\phi_k=\arctan(k_y/k_x)$
we obtain the angular dependence of $\Delta_{\rm ind}$. This is shown in 
Fig.~\ref{fig:delta0_0}~(b) for the case when the twist angle is zero.
We see that $\Delta_{\rm ind}$ is strongly anisotropic, with a $C_{3v}$ symmetry,
a reflection of the structure of the reconstructed FS, Fig.~\ref{fig:K0}~(e),~\ref{fig:K2K6}~(c).
%
%
%
%

As the twist angle $\theta$ increases $\Delta_{\rm ind}$ decreases becoming vanishing small for $\theta\gtrsim 9\degree$.
Figure~\ref{fig:delta0_0}~(c) shows $E_c(\kk)$ when $\theta=9\degree$. From this figure we see that
the location where $E_c(\kk)$ is minumum appears to correspond to the original graphene's FS for which $|\kk|=k_{F,g}$.
A closer inspection however reveals small oscillations as a function of $\phi_k$, as shown in
Fig.~\ref{fig:delta0_0}~(d) where $E_c(\kk)$ is plotted as function of $\phi_k$  
and $|\kk|$ for a small range of $|\kk|$ centered at $k_{F,g}$.

%
%

%
%

\begin{figure}[!h]
 \begin{center}
  \centering 
  \includegraphics[width=\columnwidth]{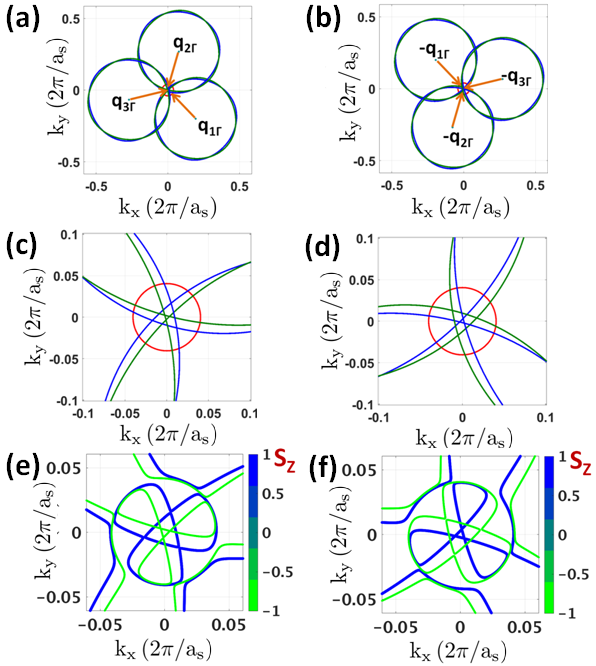}
  \caption{Fermi surfaces for $\theta=20\degree$, situation for graphene's FS overlaps with \nbse's pocket $\Gamma$. 
           Left and right panels show the results for the Dirac bands at valley $K$ and $K'$, respectively. 
           (a),~(b) FSs  for $t=0$. (c),~(d) zoom of (a) and (b), respectively. 
           (e),~(f) FSs  for $t=20$~meV.
          } 
  \label{fig:G20}
 \end{center}
\end{figure}

We now consider the case when the graphene's FS touches, in the extended BZ, the \nbse's FS pocket around the $\Gamma$ point.
Figure~\ref{fig:G20} shows the results when $\theta=20\degree$, situation for which the overlap between the graphene's FS and 
the \nbse's pocket at the $\Gamma$ point is largest. The left row show the results for the $\KK$ point, the right the ones for the $\KK'$ point.
Figure~\ref{fig:G20}~(a),~(b), show, on a fairly large scale, the configuration of the graphene's and \nbse' FSs, in the absence
of any interlayer tunneling, and the corresponding ${\qq_i}$ vectors.
Figure~\ref{fig:G20}~(c),~(d) show a zoom, at small momenta, of Fig.~\ref{fig:G20}~(a) and (b), respectively, from which we can
see that the graphene's FS and the \nbse's spin-split FS intersect at several points.
At these intersections the graphene's and \nbse's states strongly hybridize causing the FS of the system to take
the form shown in Fig.~\ref{fig:G20}~(e),~(f), for the case when $t=20$~meV, and $\Delta_\Gamma=0$.

\begin{figure}[!h]
 \begin{center}
  \centering 
  \includegraphics[width=\columnwidth]{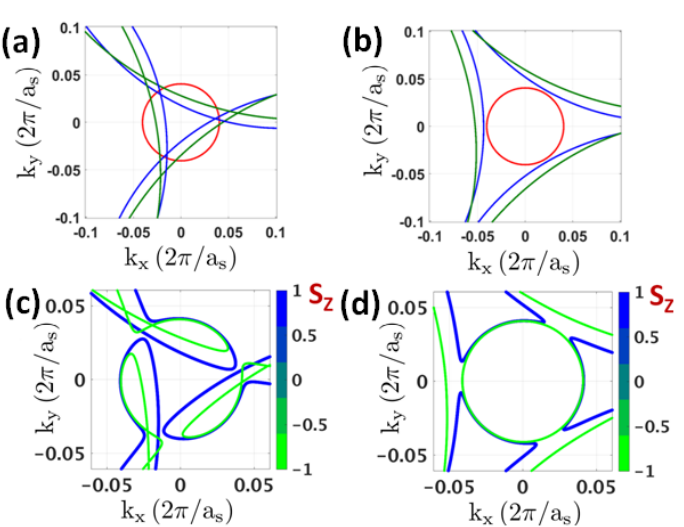}
  \caption{(a) FSs for  $\theta=18\degree$ and $t=0$.
           (b) FSs for  $\theta=16\degree$ and $t=0$.
           (c) FSs for  $\theta=18\degree$ and $t=20$~meV.
           (d) FSs for  $\theta=16\degree$ and $t=20$~meV.
         } 
  \label{fig:G16G18}
 \end{center}
\end{figure}

As $\theta$ moves away from $20\degree$ the overlap of the  graphene's and \nbse's FSs is reduced. 
For $\theta=18\degree$ the overlap is still significant, the graphene's and \nbse's FS still intersect, Fig.~\ref{fig:G16G18}~(a),
resulting in a significantly modified FS for the graphene-\nbse system, Fig.~\ref{fig:G16G18}~(c).
%
For $\theta=16\degree$ the graphene's and \nbse's FSs merely touch, Fig.~\ref{fig:G16G18}~(b). 
As a consequence the FS of the hybridized
system, for $t=20$ and \deltag=0,  is quite similar to the FS of the two isolated systems.

%

\begin{figure}[htb]
 \begin{center}
  \centering 
  \includegraphics[width=\columnwidth]{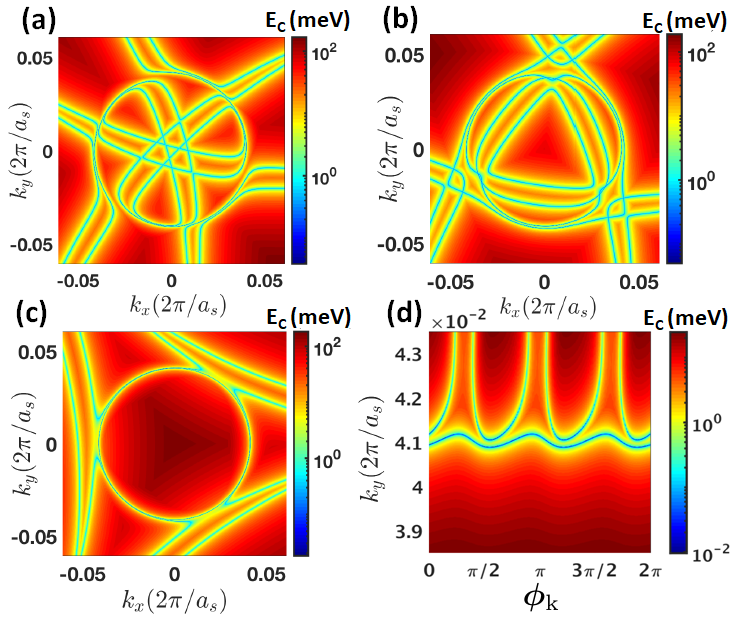}
  \caption{
           $E_c(\kk)$ for: $\theta=20\degree$, (a), $\theta=22\degree$, (b), and $\theta=16\degree$, (c).
           For  $\theta=16\degree$ the induced superconducting gap is very small.
           Panel~(d) shows the value of $E_c(\phi_k,|\kk|)$ for  $\theta=16\degree$.
          } 
  \label{fig:deltaG}
 \end{center}
\end{figure}

The superconducting gap on the \nbse's Gamma pocket induces a gap in the graphene layer when $\theta$ is around $22\degree$.
Figure~\ref{fig:deltaG}~(a)-(c) show the profile of $E_c(\kk)$ for $\theta=(20\degree,22\degree,16\degree)$, respectively.
As $\theta$ moves away from $22\degree$ $\Delta_{\rm ind}$ decrease.
Figure~\ref{fig:deltaG}~(d) show $E_c(\kk)$ as function of $\phi_k$  
and $|\kk|$ for a small range of $|\kk|$ centered at $k_{F,g}$ for the case when $\theta=16\degree$
and the original FSs of graphene and \nbse barely touch. As for the case then $\theta=9\degree$
we see that also for $\theta=16\degree$ $\Delta_{\rm ind}$ is very small and oscillates as function
of $\phi_k$ for $|\kk|\approx k_{F,g}$.

\begin{figure}[htb]
 \begin{center}
  \centering 
  \includegraphics[width=\columnwidth]{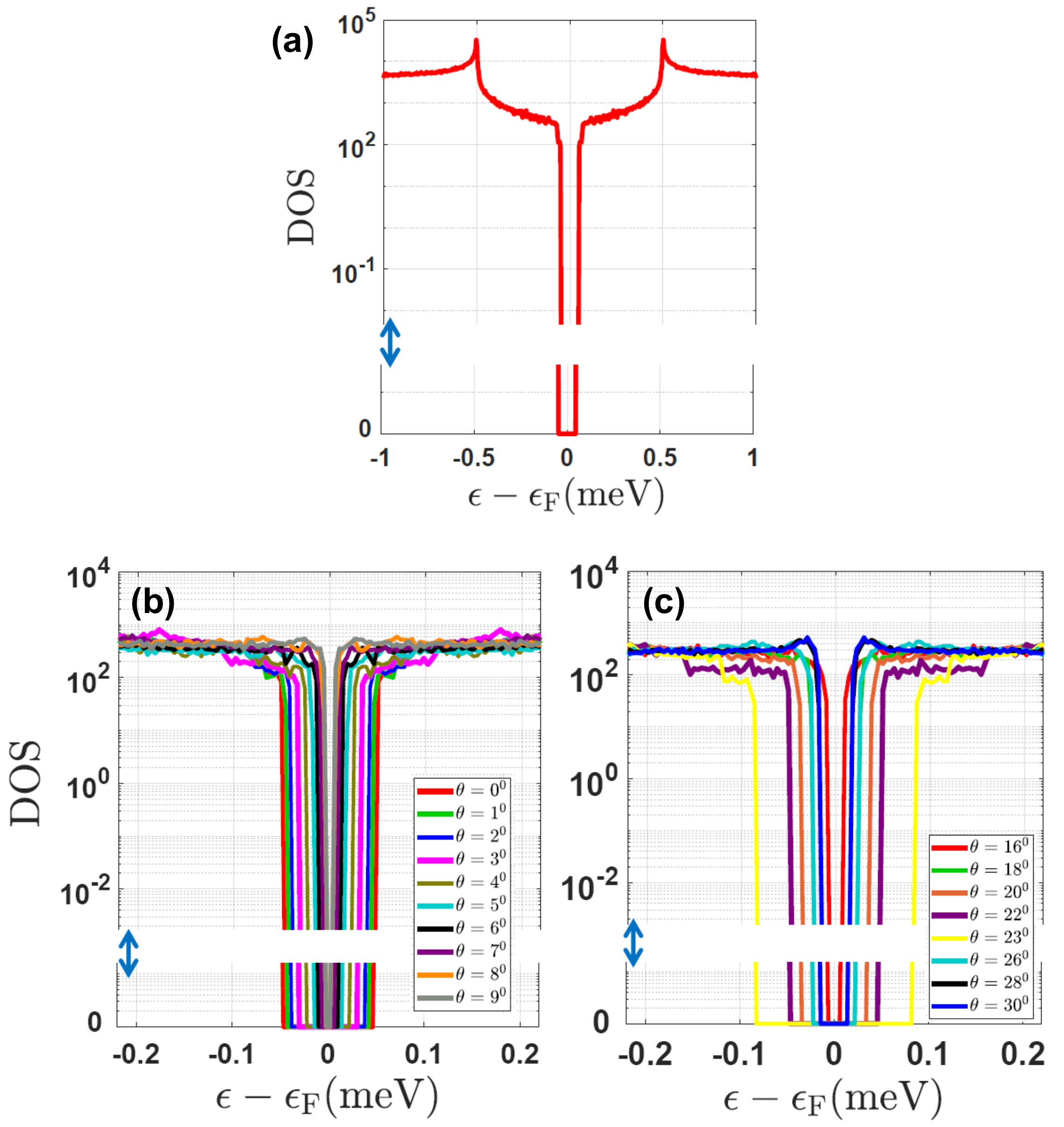}
  \caption{
           (a) Plot full DOS for graphene-\nbse heterostructure for $\theta=0.$.
           (b) Low energy zoom of panel (a), for several values of $\theta$
           for which the graphene's FS is touching \nbse $K$ point valley.
           (c) Same (b) for values of $\theta$ for which
           the graphene's FS overlaps with \nbse pocket around the $\Gamma$ point. 
          } 
  \label{fig:DOS}
 \end{center}
\end{figure}

Using tunneling experiments~\cite{steinberg2015,dvir2018} it is possible to obtain the density of states, DOS, of van der Waals
systems like graphene-\nbse. From the DOS it is then straightforward to extract the value of
the induced superconducting gap.
Figure~\ref{fig:DOS}~(a) shows the total DOS as a function of energy 
on a linear-log scale. We observe the coherence peaks corresponding to
the \nbse's superconducting gap. Below such coherence peaks the DOS remains finite,
because of the graphene's states, until the energy is equal to \deltaind. 
When the energy is equal to \deltaind the DOS rapidly goes to zero given that at that energy
also the graphene's states become gapped.
By analyzing the DOS at small energies we can find how it depends on the twist angle,
as shown in Fig.~\ref{fig:DOS}~(b),~and~(c).
Figure  \ref{fig:DOS}~(b)
shows the low energy DOS for several values
of $\theta$ close to zero, i.e., for the case when $\KK_g$ is close to $\KK_s$, and 
Figure~\ref{fig:DOS}~(b) shows it for several values
of $\theta$ close to $20\degree$, i.e., for the case when the $\KK_g$ is close to $\Gamma$ point
of \nbse's extended BZ.
%



\begin{figure}[htb]
	\begin{center}
		\centering 
		\includegraphics[width=\columnwidth]{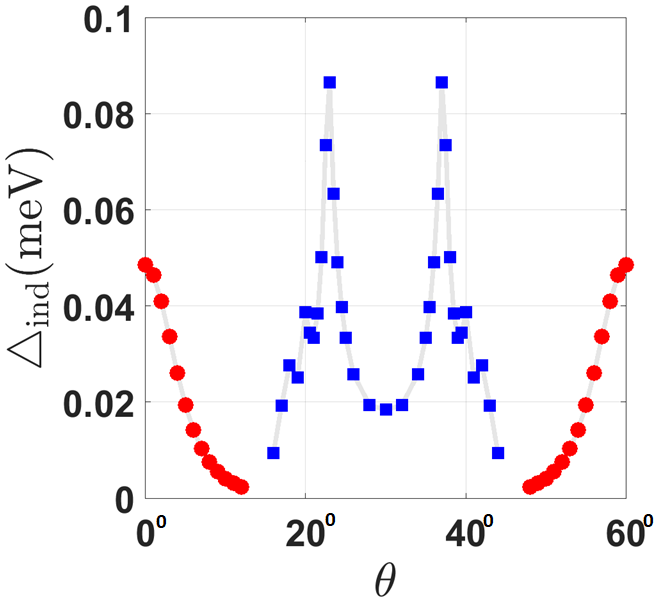}
		\caption{Induced gap \deltaind as a function of twist angle $\theta$.
		} 
		\label{fig:Delta_ind}
	\end{center}
\end{figure}

From results like the ones showed in Figs.~\ref{fig:DOS}~(b),~(c), we can extract the size of the induced superconducting gap 
and in particular its dependence on the twist angle, Fig.~\ref{fig:Delta_ind}. 
%
%
%
We see that \deltaind has a fairly sharp peak for $\theta=23\degree$ (we used a $0.5\degree$ resolution) where
it reaches the value of 0.087~meV.
This is due to the fact that for $\theta\approx 23\degree$ there is a very strong overlap
of the graphene's and \nbse Fermi surfaces.
\deltaind rapidly decrease as $\theta$ deviates from $23\degree$ and becomes an order of magnitude smaller
when $\theta=16\degree$.
$\Delta_{\rm ind}(\theta)$ has a lower and broader peak for $\theta=0$, for wich \deltaind=0.05~meV,
i.e., for the situation 
in which the graphene's FS has the maximum overlap with the \nbse $K$ pockets.
As $\theta$ increases from zero \deltaind smoothly decreases and becomes negligible for $\theta\approx 9\degree$.
Due to the symmetry of the system the behavior of $\Delta_{\rm ind}(\theta)$ has a ``mirror'' symmetry
around $\theta=30\degree$ and is periodic with period equal to $60\degree$, as exemplified by Fig.~\ref{fig:Delta_ind}.
We notice that the range of values of $\theta$ for which 
\deltaind is not vanishingly small is larger than what we can infer by simply looking at the overlaps
of the graphene's and \nbse's FSs, Fig.~\ref{fig:FS}. 
The reason is that for finite $t$ graphene's and \nbse's states that are within the energy window
$|t|$ can still hybridize resulting in a nonzero \deltaind.

Figure~\ref{fig:Delta_ind} shows that 
in a graphene-\nbse structure the superconducting gap can be strongly tuned
by varying the twist angle and that, counterintuitively, the maximum induced
gap is achieved for a value of $\theta$ for which the graphene's FS overlaps
with the $\Gamma$ pocket of \nbse  in the second BZ.

\begin{figure}[htb]
	\begin{center}
		\centering 
		\includegraphics[width=\columnwidth]{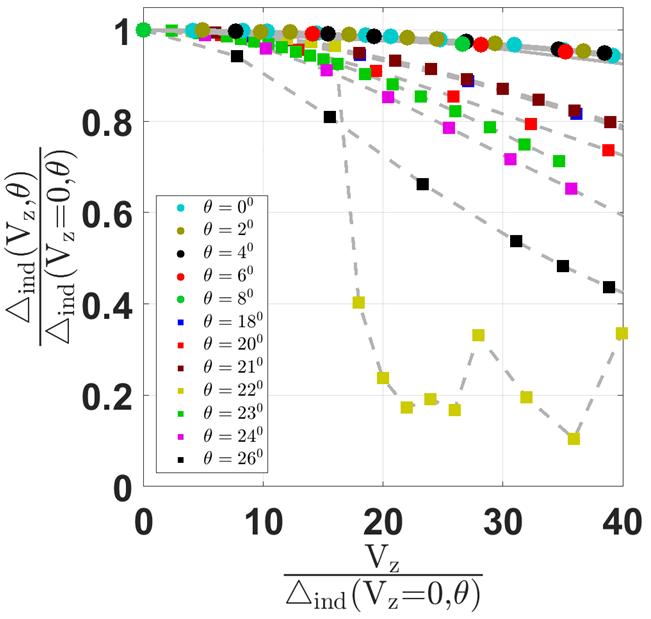}
		\caption{Figure (a): Induced gap \deltaind as a function of Zeeman field, $V_z$. 
                         The solid lines (circles) show the results for values of $\theta$ for which
                         graphene's FS overlaps with \nbse's $K$ pockets.
                         The dashed lines (squares) show the results for values of $\theta$ for which
                         graphene's FS overlaps with \nbse's $\Gamma$ pocket.
		} 
		\label{fig:zeeman_effect}
	\end{center}
\end{figure}

Due to the strong spin-orbit coupling in \nbse the in plane critical field is much larger than the 
field corresponding to the Pauli paramagnetic limit. Due to the fact that SOC is also induced
into the graphene layer via proximity effect we find that also for graphene-\nbse
heterostructures the in plane upper critical field is much larger than the Pauli paramagnetic limit.
This is shown in Fig.~\ref{fig:zeeman_effect} in which we plot the evolution of 
\deltaind in the presence of a Zeeman term $V_z$ both for values of $\theta$ corresponding
to the case when the graphene's FS overlaps \nbse's $K$ pockets (solid lines and circles),
and for values of $\theta$ corresponding
to the case when the graphene's FS overlaps \nbse's $\Gamma$ pocket
(dashed lines and squares).
We see that in both cases \deltaind remains 
finite for $V_z$ as large as 40 times the induced gap of the system at zero magnetic field.
However, it is also evident that the suppression of \deltaind due to the magnetic field
is weaker, and almost independent of $\theta$, for the case when 
graphene's FS overlaps \nbse's $K$ pockets. 
This is a consequence of the fact that in \nbse the bands' spin splitting due to SOC is
much stronger for the $K$ pockets than for the $\Gamma$ pocket.

\begin{figure}[h!!!!!tb]
	\begin{center}
		\centering 
		\includegraphics[width=\columnwidth]{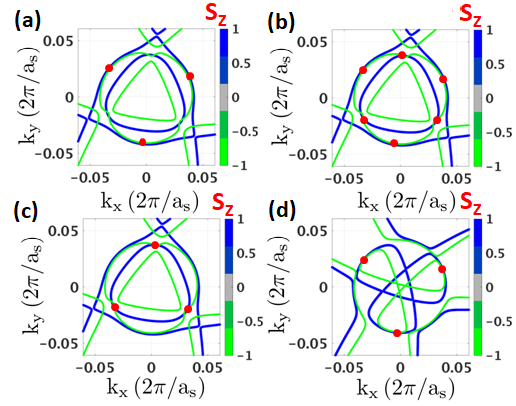}
		\caption{Location $\kk^*$ in momentum space where \deltaind is minimum:
                         (a) $\theta = 22\degree$, $V_z=0$;
                         (b) $\theta = 22\degree$, $V_z=14\Delta_{\rm ind}(V_z=0)$;
                         (c) $\theta = 22\degree$, $V_z=16\Delta_{\rm ind}(V_z=0)$;
                         (d) $\theta = 20\degree$, $V_z=0$;
                       } 
		\label{fig:minimum_gap}
	\end{center}
\end{figure}

From Fig.~\ref{fig:zeeman_effect} we notice that
for $\theta=22\degree$ the dependence 
of \deltaind on the Zeeman term deviates from the dependence  that we find
for the other values of $\theta$: 
\deltaind suddenly decreases when $V_z \approx 15\Delta_{\rm ind}(V_z=0)$, and it
exhibits oscillations for larger values of $V_z$.
The reason is that for this value of $\theta$ there are several points
in momentum space for which the induced gap is close to the minimum value and,
as shown in Figs.~\ref{fig:minimum_gap}~(a)-(c), as $V_Z$ increases the point, $\kk^*$, in momentum space where the induced gap is minimum moves.
This is in contrast to what happens for other value of $\theta$, for which the 
gap is minimum always around the same points in $k$ space, Figs.~\ref{fig:minimum_gap}~(d), regardless of the value of $V_z$.
This implies, for $\theta=22\degree$, depending on the value of $V_z$
the minimum gap will be located at points with significantly different amount of SOC-induced spin splitting of the original FSs,
and therefore different robustness against an in-plane magnetic field.
%

\section{Conclusions}

In conclusion, we have shown that, despite the large lattice mismatch between graphene's and monolayer \nbse's lattice constants,
in graphene-\nbse heterostructures graphene exhibit a significant proximity-induced superconducting gap for
a large range of stacking configurations. This is due to the fact that \nbse has large FS pockets that
overlap with the FS of graphene for most twist angles. 
Using ab-initio calculations we have obtained the amount of charge transfer between graphene and \nbse and 
estimated the strength of the interlayer tunneling. We have then obtained a continuum model 
to describe the low energy electronic structure valid in the limit of small interlayer tunneling,
condition that the ab-initio results show is satisfied. 
The continuum model takes into account both the presence of SOC and superconducting pairing in \nbse
and the fact that, depending on the twist angle, graphene's FS overlaps either with 
\nbse's FS around the $\KK$ point or the $\Gamma$ point. Using this model, and the value
of the parameters from ab-initio calculations, we find that, assuming conservatively the gap in \nbse monolayer to be equal to 0.5~mev,
and the graphene-\nbse tunneling to be 20~meV,
the maximum
induced superconducting gap in graphene is $\sim 0.09$~meV, obtained for a situation when the
graphene FS has maximum overlap the \nbse's FS around the $\Gamma$ point. 
We have shown that the superconducting 
gap induced into the graphene layer is very robust to external in plane magnetic fields: the superconducting
gap remains finite for values of the Zeeman term more than 40 times larger then the value of the induced
gap in the absence of magnetic fields.
In addition, we have shown that such robustness strongly depends on the twist angle in the sense
that if $\theta$ is such that the graphene's FS overlaps with the \nbse pockets around the $K$ points
the induced gap is much more robust to an external in-plane magnetic field
than if $\theta$ is such that the graphene's FS overlaps with the \nbse pocket around the $\Gamma$ pocket.
This is a consequence of the fact that the spin-splitting of the \nbse bands due to SOC
is much stronger at the $K$ point than at the $\Gamma$ point.

The strong dependence on the external magnetic fields of the superconducting gap induced into the graphene layer
is a reflection of the fact that graphene can be used, by simply varying the twist angle, 
as a momentum-selective probe of the electronic structure, and properties, of the substrate.
We can therefore envision that tunneling experiments on graphene-based heterostructures
could provide very useful, momentum selective, information on the gap structure of systems
with more complex gap profiles.

Considering the similarities between the Fermi surface structure of monolayer \nbse and other transition metal
dichalcogenides our results are relevant also to other graphene-TMDs heterostructures.
This also applies to the case in which, instead of a monolayer, a few atomic layers
TMD is used. Our results suggests that in general, for a large range of stacking configurations, the graphene and TMD
states, despite the large lattice mismatch, are expected to hybridize and, when the TMD is superconducting, 
induce a significant superconducting gap into the graphene layer.
It would be interesting to study how such proximity affect can affect the ground state of 
twisted-bilayer graphene systems~\cite{lu2016,Cao2018,Cao2018b,Yankowitz2019,leroy2018}.

\section{Acknowledgments}

We thank Eric Walters for helpful discussions. 
This work is supported by BSF Grant 2016320. 
YSG and ER acknowledge support from NSF CAREER grant DMR-1350663.
HS is supported by  European Research Council Starting Grant (No. 637298, TUNNEL).
ER also thanks ONR and ARO for support.
The numerical calculations have been performed on computing facilities at William \& Mary 
which were provided by contributions from the NSF, the Commonwealth of Virginia Equipment Trust Fund, and ONR.


%



%

\end{document}